\pdfoutput=1

\documentclass[runningheads]{llncs}
\usepackage{makeidx}
\usepackage{graphicx}
\usepackage{amsmath,amssymb} 
\usepackage{mathtools}
\usepackage{bm}
\usepackage{color}
\usepackage{textcomp}
\usepackage{caption}
\usepackage{cite}
\usepackage[hidelinks]{hyperref}
\usepackage[utf8]{inputenc}
\usepackage[misc]{ifsym}
\usepackage[labelformat=simple]{subcaption}
\newcommand*\diff{\mathop{}\!\mathrm{d}}

\graphicspath{{Images/}}

\begin{document}

\pagestyle{headings}
	\mainmatter

	\title{An Analysis by Synthesis Approach for Automatic Vertebral Shape Identification in Clinical QCT}

	\titlerunning{Automatic Vertebral Shape Identification in Clinical QCT}
	\authorrunning{S. Reinhold et al.}
	\author{\href{https://orcid.org/0000-0003-3117-1569}{Stefan Reinhold}\inst{1}\href{mailto:sre@informatik.uni-kiel.de}{\textsuperscript{(\Letter)}} \and 
			\href{https://orcid.org/0000-0002-5595-5205}{Timo Damm}\inst{2} \and
			Lukas Huber\inst{2} \and
			Reimer Andresen\inst{3} \and
			Reinhard Barkmann\inst{2} \and
			\href{https://orcid.org/0000-0003-3539-8955}{Claus-C. Glüer}\inst{2} \and
			\href{https://orcid.org/0000-0003-4398-1569}{Reinhard Koch}\inst{1}}

	\institute{\href{https://www.mip.informatik.uni-kiel.de/en}{Department of Computer Science, Kiel University, Kiel, Germany}\\
			   \href{mailto:sre@informatik.uni-kiel.de}{\email{sre@informatik.uni-kiel.de}} \and
			   \href{http://www.moincc.de}{Section Biomedical Imaging, Molecular Imaging North Competence Center (MOIN CC), Department of Radiology and Neuroradiology, University Medical Center Schleswig-Holstein (UKSH), Kiel University, Kiel, Germany} \and
			   \href{https://www.westkuestenklinikum.de}{Institute of Diagnostic and Interventional Radiology/Neuroradiology, 
			   Westküstenklinikum Heide, Academic Teaching Hospital of the Universities of Kiel,
			   Lübeck and Hamburg, Heide, Germany}}

	\maketitle

	\begin{abstract}
		Quantitative computed tomography (QCT) is a widely used tool for osteoporosis diagnosis and monitoring.
		The assessment of cortical markers like cortical bone mineral density (BMD) and thickness is a demanding task, mainly because of the limited spatial resolution of QCT.
		We propose a direct model based method to automatically identify the surface through the center of the cortex of human vertebra.
		We develop a statistical bone model and analyze its probability distribution after the imaging process.
		Using an as-rigid-as-possible deformation we find the cortical surface that maximizes the likelihood of our model given the input volume.
		Using the European Spine Phantom (ESP) and a high resolution \textmu CT scan of a cadaveric vertebra, we show that the proposed method is able to accurately identify the real center of cortex ex-vivo.
		To demonstrate the in-vivo applicability of our method we use manually obtained surfaces for comparison.
		
	\end{abstract}
	
	\section{Introduction}
	\label{sec:Introduction}
	
	\setcounter{footnote}{0}
	
	Osteoporosis is a systematic skeletal disease that is characterized by low bone mass and deterioration of bone microstructure resulting in high fracture risk \cite{consensus1993consensus}.
	Its high prevalence of 24\% in women beyond age of 65 makes osteoporosis a wide spread disease and a highly relevant research topic \cite{Fuchs:2017ct}.
	Quantitative computed tomography (QCT) has become a reliable tool for osteoporotic fracture risk prediction \cite{Andresen:1999bp,Guglielmi:1994bz} and monitoring \cite{gluer_comparative_2013,Genant:2016in}.
	Volumetric trabecular \emph{bone mineral density} (BMD) was identified as a good marker for bone strength. 
	Still, the high under-diagnosis rate of 84\% \cite{Smith:2004kr} indices that trabecular BMD alone is not sufficient as a bone strength marker.
	In osteoporotic patients the cortex takes the main load \cite{Rockoff:1969up}.
	Therefore, the vertebral cortical bone is a worthwhile subject of study \cite{Haidekker:1997gx,haidekker1999relationship}.
	However, the assessment of cortical markers like cortical thickness or cortical BMD is a challanging task \cite{Prevrhal:1999}:
	the spatial resolution of clinical QCT ranges from 0.3-0.5 mm in-plane and from 1-3 mm out-of-plane.
	The thickness of the cortex of a vertebral body is reported \cite{Silva:1994fg,Ritzel:1997hd} to range from 0.25 mm to 0.4 mm and is therefore clearly below the Nyquist-Frequency, resulting in tremendous overestimation of cortical thickness in clinical QCT.
	Using high resolution QCT (HR-QCT) an in-plane resolution of up to 0.15 mm and an out-of-plane resolution of about 0.3 mm can be achieved at the expense of higher radiation dose, but cortical thickness is still clearly overestimated \cite{gluer_comparative_2013}.
	
\begin{figure}[tb]
		\centering
		\includegraphics[width=0.95\textwidth]{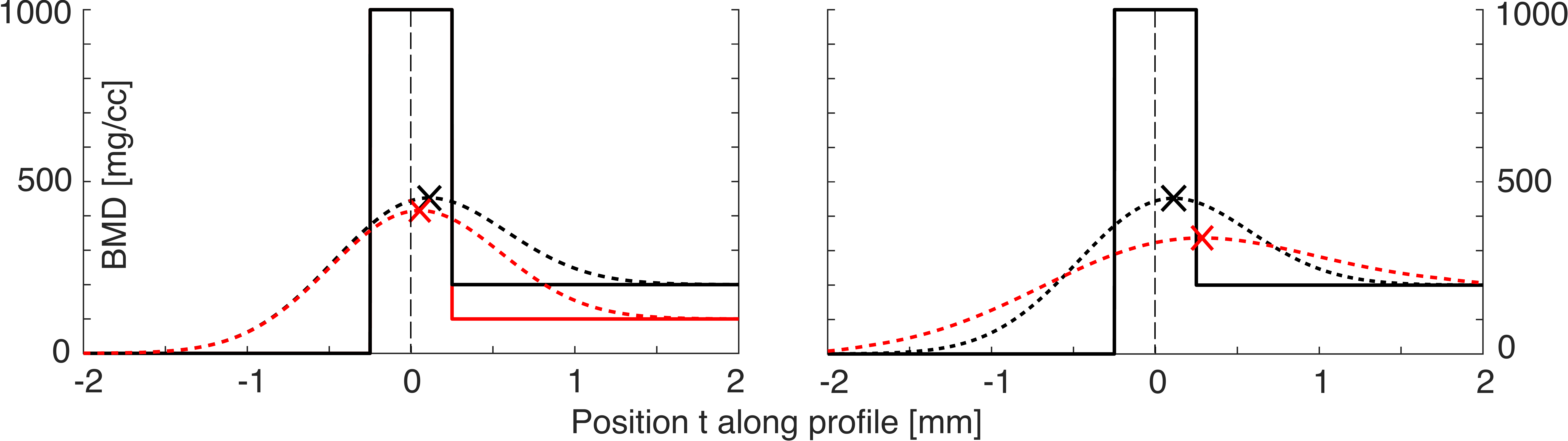}
		\caption{Cortex center shift by different trabecular BMD and resolution. Left: a simulated BMD profile trough an idealized cortex (center at $t=0$, width $0.5$ mm, mineralization $1000$ mg/cc). The trabecular BMD is $100$ mg/cc (red, left) and $200$ mg/cc (black). The dashed lines show the corresponding signals after convolution with a gaussian with $\sigma=0.5$ (black) and $\sigma = 0.8$ (red, right) simulating a clinical QCT. The crosses mark the cortical ridges (maximum intensities).}
		\label{fig:CortexCenterShift}
	\end{figure}

	In this paper we address the problem of identifying the center of the cortex of a vertebral body from clinical QCT scans without any user interaction.
	As can be seen in figure \ref{fig:CortexCenterShift} the apparent ridge of the cortical bone, i.e. the surface of maximum intensity, moves when the ratio of cortical to trabecular BMD changes.
	The same is true for different scanner resolutions \cite{Prevrhal:1999} and cortical thicknesses.
	The strength of this effect does also vary with scanner, resolution and reconstruction kernel, making results from different scanners hard to compare \cite{Giambini:2015it}.
	A direct deconvolution of the resulting image is not applicable due to low signal to noise ratio.
		
	\subsubsection{Related Work}
	
	Shape identification and segmentation of vertebral bodies has been a research topic for a long time.
	Kang et al. \cite{Kang:2003gn} use a region growing approach followed by a refinement step based on relative thresholding to distinguish soft tissue from bone.
	While the method yields accurate results for thick cortices (5 times the voxel size and above), it shows the typical overestimation for thin cortices as present in human vertebrae.
	Mastmayer et al. \cite{Mastmeyer:2005,Mastmeyer:2006kh} proposed a multi step semi-automatic segmentation method based on the Euler-Lagrange equation and local adaptive volume growing that yields promising results, but again suffers from partial volume effects for thin cortices.
	The graph cut method proposed by Aslan et al. \cite{Aslan:2009bd,Aslan:gb} shows superior performance over previous work, but since the segmentation is voxel based, voxel size, especially out-of-plane, remains the limiting factor.
	
	All of the methods mentioned yield voxel masks as the segmentation result. Therefore, the voxel size constraints the accuracy of cortex identification.
	Treece et al. \cite{Treece:2010jm} developed a mathematical model of the bone anatomy and applied a simplified model of the imaging system.
	The resulting measurement model is then fitted to the data given an initial segmentation of the target bone.
	This way they were able to provide an unbiased cortical thickness estimate down to $0.3$ mm on proximal femur.
	
	\subsubsection{Our Approach}
	We propose an analysis by synthesis (AbS) based approach (also called direct method) to accurately fit a template surface to the center of the cortical bone underlying a clinical QCT scan.
	The idea behind AbS based image analysis is not to analyze image features directly but instead  synthesize an artificial image of a parametrizable model and find the parameters for which the synthetic and the input image match best.
	This way no image derivatives are required, making the process robust to noise.
	This model based approach can, to a certain degree, compensate for loss of information in the imaging process by incorporating prior knowledge into the model.
	However, for CT a full synthesis would require a radon transform of the full model followed by a CT reconstruction.
	Since this full synthesis is very expensive computationally, we simplify the scanning process to a blur, implemented by a convolution with the point spread functions (PSF) of the system. 
	To simplify the synthesis further, we make use of \emph{sparse synthesis} \cite{Jordt:2011fr,Reinhold:2016fy} where not the whole image is synthesized, but only a sparse subset of it.
	Here, the result of the sparse synthesis is a set of one dimensional profiles orthogonal to the cortical surface, equivalent to a sparse sampling of a full synthesis\footnote{The equivalence is actually not given for the full volume. In areas where two cortices are close together, the sparse synthesis differs from the full synthesis. However, these regions are not critical for clinical routine.}.
	Our model consists of a closed genus 0 surface representing the center of the cortex.
	As in medial representations \cite{Siddiqi:1639302}, we assign a thickness and a BMD value to every point on the surface.
	The trabecular region inside the bone and the soft tissue region outside are represented by BMD distributions.
	Using a maximum a posteriori (MAP) estimation, we find the surface that maximizes the likelihood of the model parameters given measurements of the input volume.

	In chapter \ref{sec:Method} we give a detailed insight into the proposed method. We develop a statistical model of bone by extending the deterministic model of Treece et al. \cite{Treece:2010jm} in chapter \ref{sec:BoneModel}.
	The sparse synthesis yields a statistical measurement model which we derive in chapter \ref{sec:MeasurementModel}.
	In chapter \ref{sec:Optimization} we show how the MAP estimation can be carried out in a data parallel process by employing the as-rigid-as-possible (ARAP \cite{Sorkine:2007cm}) deformation scheme.
	In chapter \ref{sec:Experiments}, we evaluate the accuracy of our method by using the European Spine Phantom (ESP, \cite{Kalender:1995gv}).
	In addition, we further evaluate the accuracy using a \textmu CT scan of a cadaveric human vertebra. To show the applicability of our method to clinical data we compare our results on 100 in-vivo QCT scans with manually obtained annotations.
	Finally, in chapter \ref{sec:Conclusion} we conclude this article and discuss future work.
		
	\section{Method}
	\label{sec:Method}
	The input to our method is a calibrated\footnote{Houndsfield units are converted to bone mineral equivalents using known densities of a calibration phantom which is simultaneously scanned with the patient.} QCT scan of a vertebra, a pre-estimated statistical measurement model and a labeled three-dimensional sketch of the target bone as depicted in figure \ref{fig:Template}.
	Each label of the template corresponds to a differently parameterized statistical bone model.
	The output is a triangle mesh representing the unbiased surface of the cortex center.

	\subsection{Statistical Bone Model}
	\label{sec:BoneModel}
	
\begin{figure}
		\centering
		\begin{subfigure}[b]{0.38\textwidth}
			\includegraphics[width=\textwidth]{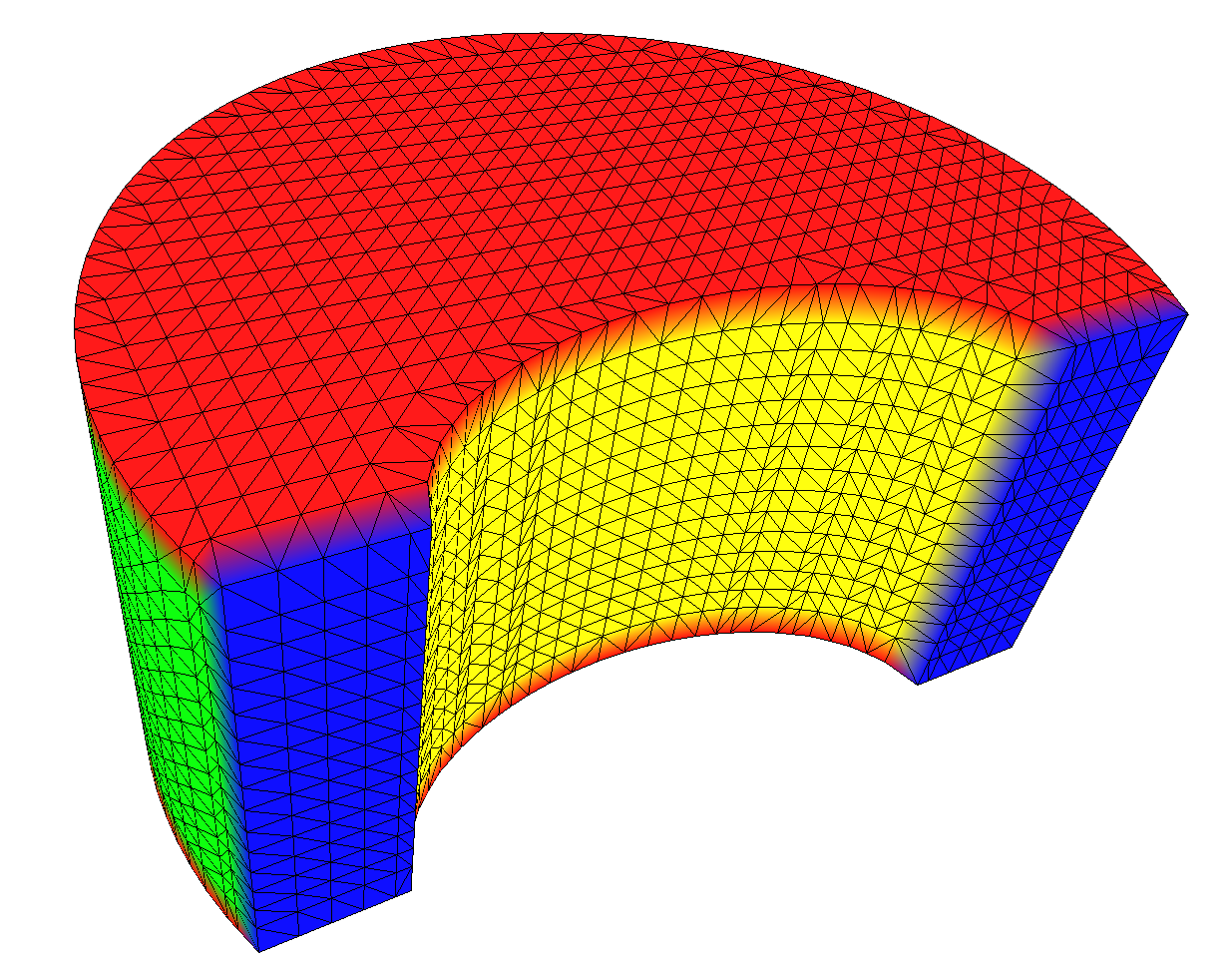}
			\caption{Labeled Template Mesh}
			\label{fig:Template}
		\end{subfigure}\hfill%
  		\begin{subfigure}[b]{0.48\textwidth}
    		\includegraphics[width=\textwidth]{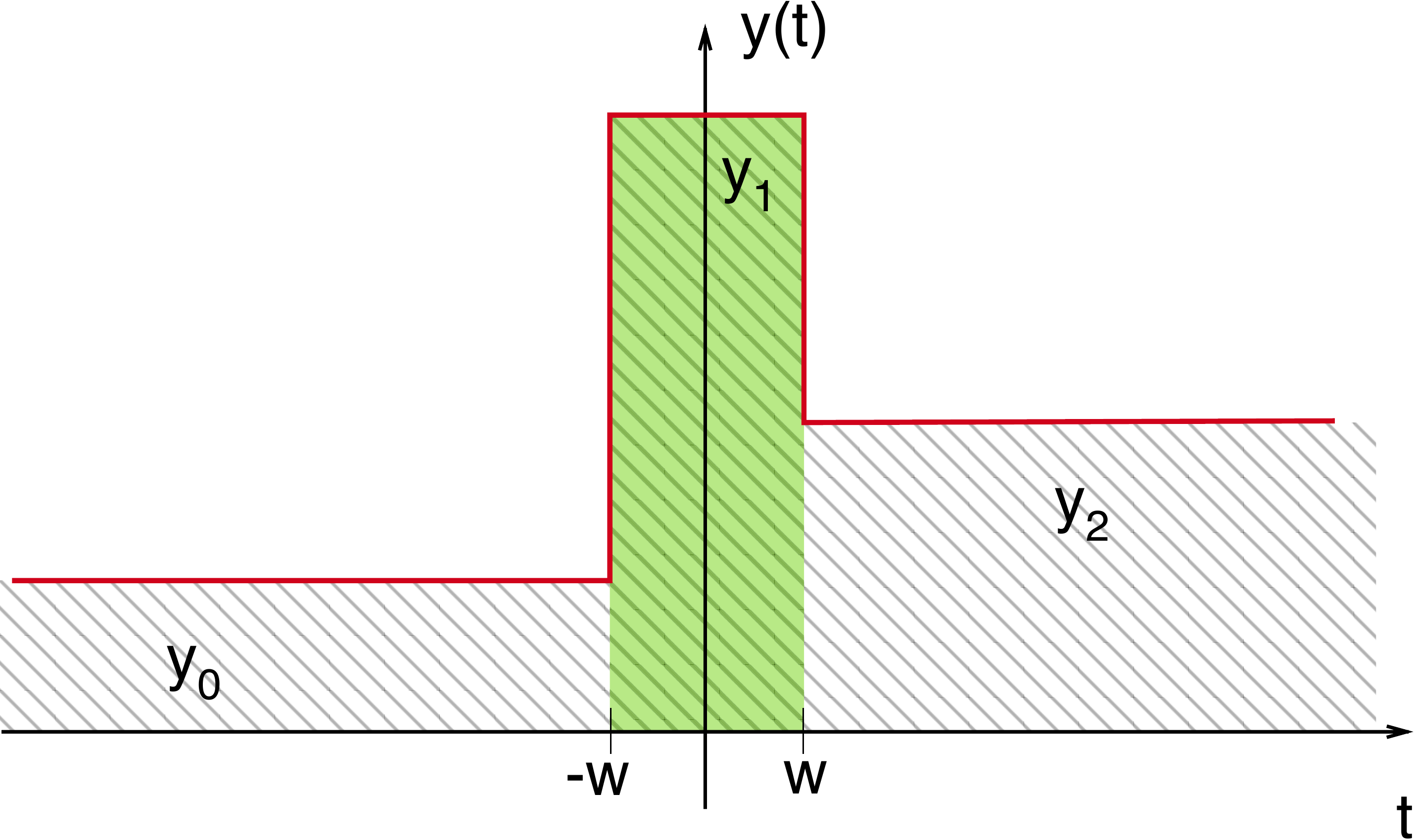}
    		\caption{Realization of Bone Model}
    		\label{fig:BoneModel}
  		\end{subfigure}
  		\caption{(a) Labeled vertebra template used in the proposed method. Different colors correspond to different regions with distinct bone models: vertical cortex (green), endplates (red), foramen (yellow), cut pedicles (blue). (b) Realization $y(t)$ of random process $Y(t)$, see eq. \ref{eq:BoneModel}. The cortex is depicted in green.}
	\end{figure}

	We model the surface of the cortex center as a closed genus 0 triangle mesh $\mathcal{S}$ with $N$ vertices.
	Its piecewise linear embedding is given by the vertex positions $V \subseteq \mathbb{R}^3$.
	When we look at a one dimensional profile orthogonal to the surface at any point on $\mathcal{S}$ (figure \ref{fig:BoneModel}), then the BMD graph is a piecewise constant function of the signed distance $t$ to the cortex center.
	By modeling the three density levels (soft tissue density, cortical BMD, trabecular BMD) as gaussian random variables $Y_i \sim \mathcal{N}(\mu_{Y_i}, \sigma_{Y_i}^2)\,, i=0,1,2$ and by modeling the cortical thickness\footnote{The cortical thickness is $2 \cdot W$.} as a random variable $W$  with\footnote{$W$ is a non-negative size quantity which is commonly modeled as a log-normal distribution.} $\log W \sim \mathcal{N}(\mu_W, \sigma_W^2)$, we obtain a stochastic process $Y(t)$ of the profile:
	
	\begin{equation}
		\label{eq:BoneModel}
		Y(t) = Y_0 + (Y_1 - Y_0) \cdot H(t + W) + (Y_2 - Y_1) \cdot H(t - W),
	\end{equation}
	where $H$ denotes the Heaviside step function.
	Note that \eqref{eq:BoneModel} takes a similar form as \cite[eq. 1]{Treece:2010jm} with all unknowns replaced by random variables.
	
	Since the density distribution of the soft tissue is not equal everywhere, for example the intervertebral discs have higher density than muscles, we use differently parameterized random processes for different sections of the bone.
	Figure \ref{fig:Template} depicts the three different regions used in our model: vertical cortex (green), endplates (red), foramen (yellow). The "cut pedicles" region (blue in fig. \ref{fig:Template}) is not used in the model.
		
	\subsection{Measurement Model}
	\label{sec:MeasurementModel}
	
	The next step in the AbS framework is the synthesis step where the imaging system is simulated by applying the in-plane and out-of-plane PSF to the bone model.
	By sampling the resulting image one can then acquire observable profiles.
	However, with sparse synthesis we are able to simplify the process by not performing the convolutions in the global coordinate system, but in the local system of the profile.
	
	\subsubsection{Combined PSF}
	The slice sensitivity profile (SSP) of a spiral CT can be approximated by the convolution of the rectangular profile of helical CT by the triangular table movement function \cite{Kalender:1994ft}.
	The width of the rectangular profile is determined by the collimation, while the width of the movement function is determined by the table feed.
	We therefore approximate the out-of-plane PSF (perpendicular to a CT slice) for pitch factor of 1 with
	\begin{equation}
		g_{O,h}(z) \coloneqq \tilde{g}_{O}\left(\frac{z}{h}\right),
	\end{equation}
	where $\tilde{g}_{O}(z) \coloneqq \mathrm{\Pi}(z) * \mathrm{\Lambda}(2z)$ and $h$ is the slice width.
	Like \cite{Treece:2010jm} we approximate the in-plane PSF with a rotational invariant gaussian of width $\sigma$:
	\begin{equation}
		g_{I}(x,y) \coloneqq \frac{1}{\sqrt{2\pi}\sigma} \exp \left(-\frac{x^2+y^2}{2\sigma^2} \right).
	\end{equation}
	
\begin{figure}[bt]
	  \centering
	  \includegraphics[width=0.9\textwidth]{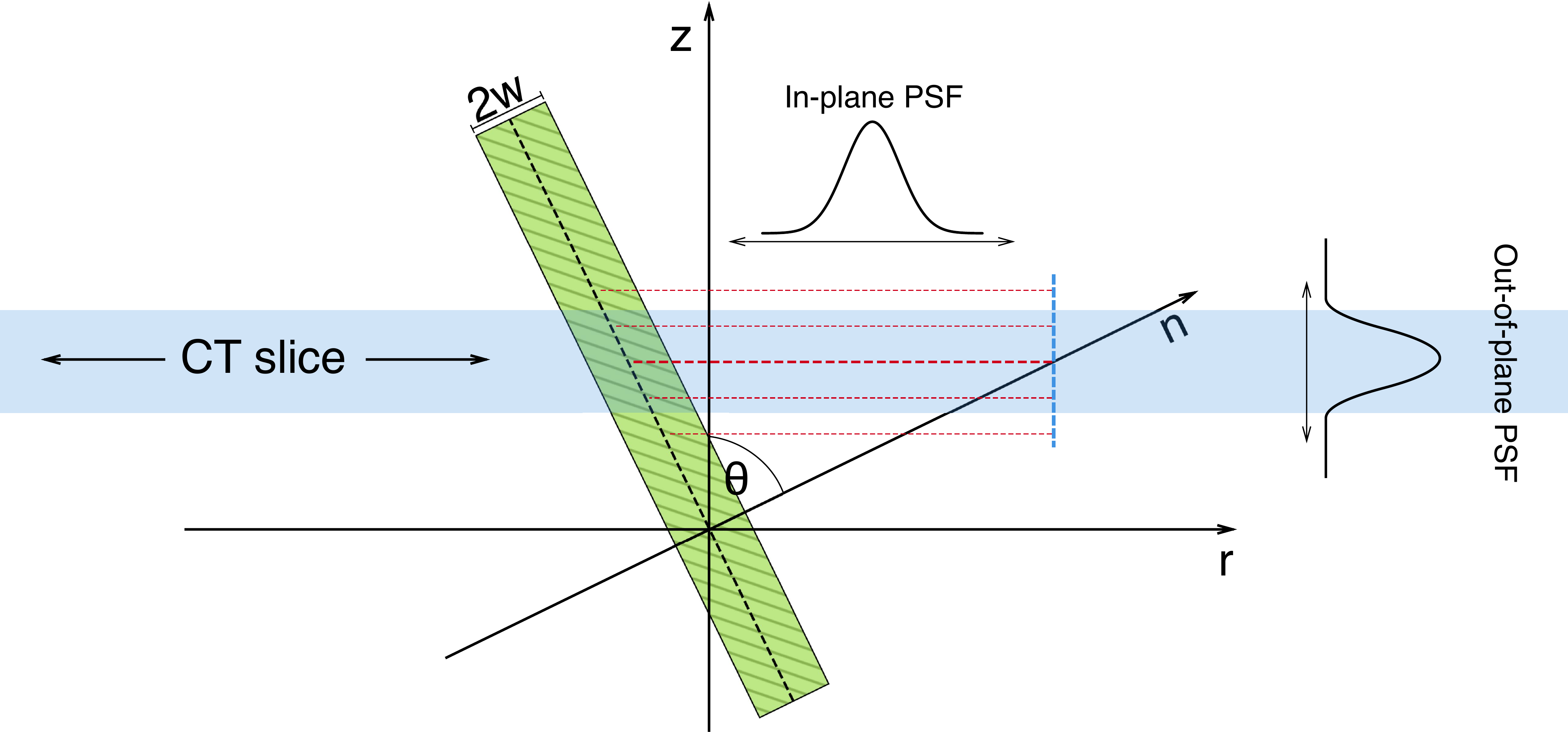}
	  \caption{Schematic view of an orthogonal cut through a cortex segment of width $2w$ (shaded green). The resulting spatial density is sampled along the surface normal $\vec{n}$ at an angle $\theta$ with the z-axis after convolution with the in-plane PSF along r and with the out-of-plane PSF along z.}
	  \label{fig:MeasurementModel}
	\end{figure}

	If we slice through the cortex at $\vec{x} \in \mathcal{S}$ with a plane defined by the z-axis and the surface normal $\vec{n}$ at $\vec{x}$, we can define a local coordinate system centered on $\vec{x}$ with the z-axis as the ordinate and the projection of $\vec{n}$ onto the x-y-plane as the abscissa (r-axis) as depicted in figure \ref{fig:MeasurementModel}.
	Since the profile now lies inside r-z-plane and because $g_I$ is rotational invariant, we can obtain the measurement process $\tilde Z(r, z)$ by convolving $\tilde Y(r, z) \coloneqq Y(r \sin \theta + z \cos \theta)$ with $g_{O,h}$ along the z-axis and with $g_I$ along the r-axis:

	\begin{eqnarray}
		\tilde Z(r, z) &=& \int \int \tilde Y(r - \tau, z - \lambda) \cdot g_I(\tau) \cdot g_{O,h}(\lambda) \diff \tau \diff \lambda \\
		&=& \int \int Y\left(t - (\tau \sin \theta + \lambda \cos \theta)\right) \cdot g_I(\tau) \cdot g_{O,h}(\lambda) \diff \tau \diff \lambda,\nonumber
	\end{eqnarray}
	where $t \coloneqq r \sin \theta + z \cos \theta$ and $\theta$ is the angle between $\vec{n}$ and the z-axis.
	By substituting $\phi \coloneqq \tau \sin \theta$ and $\psi \coloneqq \lambda \cos \theta$ we get
	\begin{eqnarray}
		\label{eq:CombinedPSF}
		\tilde Z(r, z) &=& \int \int Y(t - \phi - \psi) \cdot g_I\left(\frac{\phi}{\sin \theta}\right) \cdot g_{O,h}\left(\frac{\psi}{\cos \theta}\right) \cdot \frac{1}{\sin \theta \cos \theta} \diff \phi \diff \psi\nonumber \\
		&=& Y(t) * \left\{\frac{1}{\sin \theta}\cdot g_I\left(\frac{t}{\sin \theta}\right) * \frac{1}{\cos \theta} \cdot g_{O,h}\left(\frac{t}{\cos \theta}\right) \right\}\\
		&=& Y(t) * g_{\theta}(t).\nonumber
	\end{eqnarray}
	Hence, we can simplify the two dimensional convolution by a single convolution of the one-dimensional process from \eqref{eq:BoneModel} with an angle dependent PSF $g_\theta$.
	
	Although, we used a gaussian for the in-plane PSF in the derivation of the combined PSF, we note that any symmetric square integrable PSF can be used here.
	Ohkubo et al. \cite{ohkubo2009determination} determined the PSFs for several reconstruction kernels.
	Based on their measurements one can observe that a gaussian is a good approximation for smooth kernels like the Siemens B40, but for sharper kernels like the Siemens B80 it is not:
	the B80 amplifies some higher frequencies in a narrow band to enhance edges while the gaussian damps all high frequencies.

	\subsubsection{Stochastic Measurement Process}

	Let $G_{\theta} \coloneqq \int g_{\theta}(t) \diff t$ be the primitive function of the angle dependent PSF $g_{\theta}$ from \eqref{eq:CombinedPSF}.
	Since $g_{\theta}$ has finite energy, we know that $\displaystyle \lim_{t \rightarrow -\infty}G_{\theta}(t) = 0$ and $\displaystyle \lim_{t \rightarrow +\infty} G_{\theta}(t) = 1$.
	With this we can define the optimal (i.e. noise free) stochastic measurement process $Z(t)$ as
	\begin{eqnarray}
		\label{eq:MeasurementProcess}
		Z(t) &\coloneqq& Y(t) * g_{\theta}(t) \\
		&=& Y_0 + (Y_1 - Y_0) \cdot G_{\theta}(t+W) + (Y_2 - Y_1) \cdot G_{\theta}(t-W).\nonumber
	\end{eqnarray}
	
	Since $Y_0, Y_1, Y_2$ are stochastically independent normally distributed random variables the conditional stochastic process $\zeta(t) \coloneqq Z(t)|W$ is gaussian with mean and variance function given as:
	\begin{eqnarray}
		\mu_{\zeta}(t|w) &=& \mu_0 + (\mu_{Y_1} - \mu_{Y_0}) \cdot G_{\theta}(t+w) + (\mu_{Y_2} - \mu_{Y_1}) \cdot G_{\theta}(t-w),\\
		\sigma_{\zeta}^2(t|w) &=& \sigma_{Y_0}^2(1 - G_{\theta}(t+w))^2 + \sigma_{Y_1}^2(G_{\theta}(t + w) - G_{\theta}(t - w))^2 + \nonumber\\
		&&\sigma_{Y_2}^2 G_{\theta}^2(t-w). \nonumber
	\end{eqnarray}
	Thus the probability density function (PDF) $f_{\zeta}$ of $\zeta(t)$ is given by 
	\begin{equation}
		\label{Eq:ConditionalDensity}
		f_{\zeta}(z,t|w) = N\left(z\,\middle| \,\mu_{\zeta}(t|w), \sigma_{\zeta}^2(t|w)\right).	
	\end{equation}
	With the joint probability density function of $Z(t)$ and $W$ 
	\begin{eqnarray}
		\label{Eq:JointDensity}
		f_{Z,W}(z,w,t) &=& f_\zeta(z,t|w)f_W(w), \text{ with}\\
		f_W(w) &=& \frac{1}{w}\cdot\frac{1}{\sqrt{2\pi}\sigma_W}\exp \left(-\frac{(\ln w - \mu_W)^2}{2\sigma_W^2}\right), \nonumber
	\end{eqnarray}
	the probability density function $f_Z$ of $Z(t)$ is obtained by marginalizing out $W$:
	\begin{equation}
		\label{eq:PDF_Z}
		f_{Z}(z,t) = \int f_{\zeta}(z,t|w) f_W(w) \diff w.
	\end{equation}

	\subsection{Deformation Model and Optimization}
	\label{sec:Optimization}
	
	Given a profile through a point $\vec{x} \in \mathcal{S}$ of an input volume $I$, we can express the degree of conformity of the profile with a synthetic profile by means of the likelihood of the measurement model.
	The optimal surface $\mathcal{S}'$ is then defined by the linear embedding $V'\subset \mathbb{R}^3$ for which the likelihood of all profiles through $\mathcal{S}'$ are maximal.
	However, this is an ill posed problem, because it is not guaranteed that the maximizer yields a realistic shape.
	By constraining the deformation between an initial surface $\mathcal{S}$ and $\mathcal{S}'$ to be as rigid as possible, shape degeneration can be avoided and the problem becomes well formed.
	
	We modify the as-rigid-as-possible (ARAP) energy term from \cite{Sorkine:2007cm} to act as a shape prior in the MAP framework as follows:
	\begin{equation}
		\label{eq:ARAPEnergy}
		E(\hat{\mathcal{S}}) = \min_{R_i \in SO(3)} \left\{ \sigma_E^{-2} \displaystyle\sum_{i=1}^{N}\gamma_i\sum_{j\in\mathcal{N}(i)} \omega_{ij} \|(\hat{\vec{x}}_i - \hat{\vec{x}}_j) - \vec{R}_i ( \vec{x}_i - \vec{x}_j )\|_2^2 \right\},
	\end{equation}
	where $\omega_{ij}$ are the cotangent weights of edge $(i, j)$ \cite{Meyer:2003ed}, $\gamma_i$ are per-vertex weights and $\mathcal{N}(i)$ contains all vertices of the 1-ring of vertex $i$.
	$\vec{R}_i \in SO(3)$ are per-vertex rotation matrices adding additional $3N$ degrees of freedom to the deformation.
	$E(\hat{\mathcal{S}})$ can be seen as a measure of how far the deformation is from an isometry \cite{Chao:2010ex}. With the scale $\sigma_E^{-2}$ equation \eqref{eq:ARAPEnergy} is equivalent (up to constants) to the negative log likelihood function of a gaussian random variable $E$ modeling the ARAP energy of $\mathcal{S}'$.

	Now, let $\vec{t} = (t_i)_{1,\dots ,2K+1}$ be a vector of $2K+1$ samples from $[-t_0, t_0]\subset\mathbb{R}$ with a sampling interval $\delta_t$.
	For all $i=1, \dots, N$ let $\hat{\vec{l}}_i \coloneqq  \{\hat{\vec{x_i}} + t_j \cdot \hat{\vec{n}}_i\, |\, j=1, \dots, 2K+1\}$ be a vector of discrete samples of a profile through position $\hat{\vec{x}}_i \in \hat{V}$ and let $\hat{\vec{\rho}}_i \coloneqq I(\hat{\vec{l}}_i)$ be the densities obtained by sampling the interpolated\footnote{In our implementation we use trilinear interpolation for performance reasons.} input volume $I$ at positions $\hat{\vec{l}}_i$.
	The log likelihood $\mathcal{L}(\hat{\vec{x}}_i|\hat{\vec{\rho}}_i)$ of $\hat{\vec{x}}_i$ given measurements $\hat{\vec{\rho}}_i$ is given with \eqref{eq:PDF_Z} by\footnote{To keep the derivation simple, we assume independence here. In general the samples are correlated by the interpolation method.}
	\begin{equation}
		\label{eq:MeasurementLikelihood}
		\mathcal{L}_Z(\hat{\vec{x}}_i | \hat{\vec{\rho}}_i) = \sum_{j=1}^{2K+1} \log f_Z(\hat{\rho}_{ij}, t_j).
	\end{equation}
	The posterior log likelihood is then proportional to
	\begin{equation}
		\mathcal{L}_{E|Z}(\hat{\mathcal{S}}_i|\hat{\vec{\rho}}_i) \propto \mathcal{L}_Z(\hat{\vec{x}}_i |\hat{\vec{\rho}}_i) - E(\hat{\mathcal{S}}) - \log \int f_{Z|E}(\hat{\vec{\rho}}_i | \hat{\mathcal{S}}) \diff \hat{\mathcal{S}}.
		\label{eq:DirectPosteriorLikelihood}
	\end{equation}
	The direct optimization of \eqref{eq:DirectPosteriorLikelihood} is analytically intractable and numerical optimization would either require the computation of image gradients, which we are tying to avoid, or gradient less optimization methods which do not perform very well on high dimensional optimization problems.
	We therefore split the optimization into two sub problems that can be efficiently solved:
	first the optimal displacement along the surface normal for each profile is estimated.
	Afterwards, the surface is optimized to match the displaced positions under ARAP constraints.
	This two step optimization scheme is iterated until convergence.
	
	\subsubsection{Optimal Displacements}
	\label{sec:OptimalDisplacements}
	We first modify equation \eqref{eq:MeasurementLikelihood} by adding a latent variable $s_i \sim \mathcal{N}(0, \sigma_{s_i}^2)$ as follows:
	\begin{equation}
		\mathcal{L}_{Z|s_i}(\vec{x}_i | \vec{\rho}_i, s_i) = \mathcal{L}_Z(\vec{x}_i + s_i \vec{n}_i | \vec{\rho}_i).
	\end{equation}
	Using MAP estimation we can find the optimal displacement $\hat{s}_i$ by maximizing the posterior likelihood
	\begin{eqnarray}
		\label{eq:DisplacementPosterior}
		&&\mathcal{L}_{s_i|Z}(s_i|\vec{x}_i, \vec{\rho}_i) =\\
		&&\mathcal{L}_{Z|s_i}(\vec{x}_i | \vec{\rho}_i, s_i) + \mathcal{L}_{s_i}(s_i) - \log \int f_{Z|s_i}(\vec{\rho}_i | \vec{t}_i + \tau) f_{s_i}(\tau) \diff \tau.\nonumber
	\end{eqnarray}
	In our implementation we pre-evaluate $f_Z$ for discrete samples of $t, \theta$ and $z$ to a three dimensional histogram and approximate $f_Z$ by histogram lookup with trilinear interpolation.
	This way, equation \eqref{eq:DisplacementPosterior} can be maximized very efficiently using data parallel exhaustive search along $s_i$ at discrete steps.
	
	\subsubsection{Surface Fitting}
	Now that the optimal displacements are known, the surface needs to be fitted to the displaced positions $\vec{y}_i \coloneqq \vec{x}_i + \hat{s}_i \vec{n}_i$ under ARAP constraints.
	We therefore minimize the following energy term:		
	\begin{equation}
		E_{\text{shape}}(\hat{\mathcal{S}}) = \sum_{i=1}^N \gamma_i \left( (\hat{\vec{x}}_i - \vec{y}_i)^T \vec{n_i} \right)^2 + E(\hat{\mathcal{S}}),
		\label{eq:EShape}
	\end{equation}
	where the first term addresses the point to plane distances and the second term the ARAP constraints.
	There might be profiles $\vec{\rho}_i$ where no cortex can be observed.
	Since the estimated displacements from those profiles are meaningless, we use the per-vertex weights $\gamma_i$ to down-weight those estimates by using the posterior PDF: $\gamma_i \coloneqq f_{s_i|Z}(\hat{s}_i|\vec{x}_i, \vec{\rho}_i)$.
	
	Like in \cite{Sorkine:2007cm} we use an alternating iterative optimization scheme to optimize \eqref{eq:EShape}: we first keep the rotations $\vec{R}_i$ fixed and optimize for the positions $\hat{\vec{x}}_i$ and then optimize for the rotations while keeping the positions fixed.
	For fixed positions, the optimal rotations can be found by SVD (refer to \cite[eqs. 5,6]{Sorkine:2007cm} for details).
	The optimal positions $\hat{\vec{x}}_i$ can be found by setting the partial derivatives of $E_{\text{shape}}$ to zero, which results in the following sparse linear system of equations:
	\begin{equation}
		\left( 2 \sigma_E^{-2} L \otimes I_{3} + B\right) \vec{x} = 2\sigma_E^{-2}\vec{c} + \vec{d},
		\label{eq:SparseLSE}
	\end{equation}
	where $\vec{x} = \left(\vec{x}_1^T, \dots, \vec{x}_N^T\right)^T$, $L$ is the Laplacian matrix of $\mathcal{S}$, $B$ is a block diagonal matrix with entries $(\gamma_i \vec{n}_i \vec{n}^T)_{ii}$ and $\otimes$ denotes the Kronecker product. The vector $\vec{c}$ contains the ARAP constraints as a concatenations of vectors $\vec{c}_i = \sum_j L_{ij} \frac{( \vec{R}_i + \vec{R}_j )}{2} (\vec{x}_i - \vec{x}_j)$ and $\vec{d}$ consist of concatenations of vectors $\vec{d}_i = \gamma_i \vec{n}_i \vec{n}_i^T (\vec{x}_i + \hat{s}_i \vec{n}_i)$.
	Equation \eqref{eq:SparseLSE} can be efficiently solved using a preconditioned conjugate gradient solver (PCG).
	The alternating optimization scheme is iterated until convergence.

	\section{Experiments and Results}
	\label{sec:Experiments}
	
	We implemented the proposed method using MATLAB and C++.
	The measurement model  requires the slice spacing $h$ and the width $\sigma$ of the in-plane PSF as input parameters.
	The value of $\sigma$ can be easily estimated from phantom scans\footnote{Since the calibration phantom is present in all QCT scans, no separate scanning process is required.}.
	The parameters of the BMD priors $Y_i$, can be estimated from single measurements or prior knowledge.
	The parameters of the width prior $W$ are set to reflect the range of reported cortical thicknesses.
	For each scanner configuration $(h, \sigma)$ we pre-computed $f_Z$ for discrete samples from $(t, \theta, z) \in [-2, 2] \times [0^{\circ}, 90^{\circ}] \times [-1000, 2000]$ (41, 91 and 3001 samples, respectively) data parallel on a GPGPU and saved the result as a 3D histogram as described in section \ref{sec:OptimalDisplacements}.
	We set $\sigma_{s_i}=2$ and $\sigma_E=2$ for all experiments.
	The optimization usually provides good results after a few iterations and takes about $1$ to $5$ minutes to complete on our system (Intel\textsuperscript{\textregistered} Core\textsuperscript{\texttrademark} i7-4790 CPU @ 3.60GHz, 4 Cores), depending on the image size and number of iterations\footnote{In our prototype implementation, the optimization does not utilize the GPU, yet, but we note that all operations can be easily ported to the GPU.}.
	
	\subsection{Ex-Vivo}
	\label{sec:ExVivoExperiments}
	
	\begin{table}[bt]
	
		\centering
		\caption{Results of the phantom experiment. Given are for each of the three ESP vertebrae the radii and heights of the estimated surface. For both variables the mean, the standard deviation (SD) and the difference to the ground truth values (Diff) are shown. All measures are in mm. N denotes the number of samples on each surface (vertical cortex for radius, endplates for height).}
		\label{tab:ESPResults}
	
		\begin{tabular*}{\textwidth}{l @{\extracolsep{\fill}} rrrrc|rrrr}
			\hline
			&\multicolumn{4}{c}{\bfseries Radius}& &\multicolumn{4}{c}{\bfseries Height}\\
			Vertebra & N & Mean & Diff & SD & & N & Mean & Diff & SD\\
		  	\hline
			Low & 10072 & 17.73 & -0.02 & 0.14 & & 6101 & 23.95 & -0.05 & 0.28 \\ 
		  	Medium & 9897 & 17.52 & 0.02 & 0.16 & & 6239 & 23.64 & -0.36 & 0.42 \\ 
  			High & 9602 & 17.23 & -0.02 & 0.16 & & 23306 & 23.00 & 0.00 & 0.30 \\
   			\hline
		\end{tabular*}
		
	\end{table}
	
	To evaluate the accuracy of the proposed method, we scanned the ESP with a clinical CT scanner using a low dose protocol (Siemens Somatom 64, 120 kV, 80 mAs, kernel B40s) at an in-plane resolution of 0.4 mm and a slice spacing of 1 mm.
	The ESP consists of three geometrical phantom vertebra (low, medium and high), each with different wall thicknesses and densities.
	The proposed method was used to acquire the cortical surfaces $\mathcal{S}_i$ of the three phantom vertebrae.
	To gain a dense surface, we used random mesh sampling of $\mathcal{S}_i$ using the method described in \cite{cignoni1998metro}. For the accuracy evaluation of the vertical cortex, we first fit a cylinder to the point cloud belonging to the vertical cortex and then computed the radius for each point separately.
	Since the vertebral bodies of the ESP have a diameter of 36 mm, the optimal radii of the cylindrical surfaces through the cortex centers are $17.75$ mm ($0.5$ mm wall), $17.5$ mm ($1$ mm wall) or $17.25$ mm ($1.5$ mm wall), for the low, medium and high vertebra, respectively.
	To evaluate the accuracy of the endplate surface we fit two planes to our point cloud: one for the upper endplate and one for the lower endplate.
	Hereafter, the point to plane distance of each sample to its opposite plane was computed.
	All bodies of the phantom have a height of 25 mm, so the optimal distances are 24 mm for the low and medium vertebra ($1$ mm wall) and 23 mm ($2$ mm wall) for the high vertebra.
	
	Table \ref{tab:ESPResults} summarizes the results of the ESP experiment. For all vertebra levels, the difference between the mean radius of the estimated surface and the 	ground truth radius is near zero. The standard deviation is below one half of the in-plane resolution. The same applies for the estimated heights, noting that the out-of-plane resolution is $2.5$-times lower than the in-plane resolution.
	
\begin{figure}[bt]
		\centering
   		\includegraphics[keepaspectratio=true, width=0.95\textwidth]{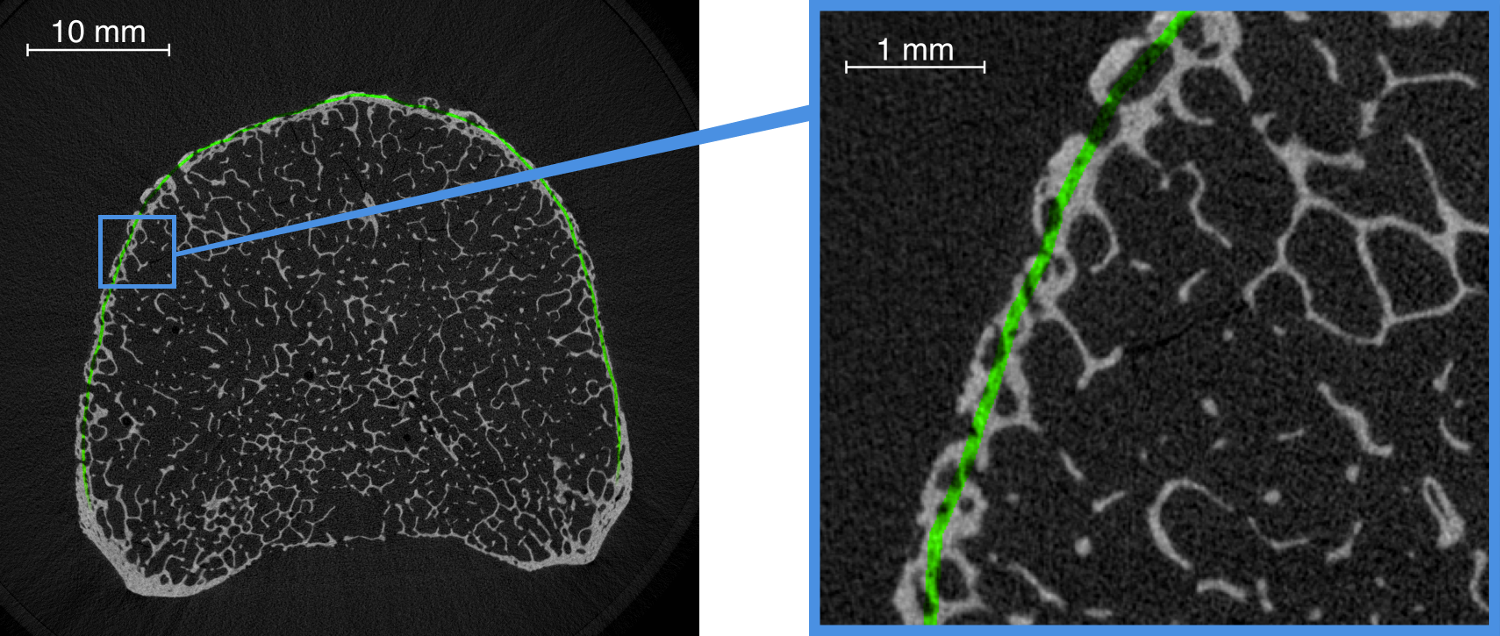}
  		\caption{Left: central slice of the embedded vertebra acquired with \textmu CT. Right: zoom to a region where cortex definition is problematic. The green contour shows the cortex center as estimated by the proposed method.}
  		\label{fig:muCTScan}
	\end{figure}

	Since the shape of the ESP is very simple, we used a cadaveric vertebra embedded in resin for a more realistic reference.
	The embedded vertebra was scanned with a \textmu CT system (SCANCO Medical, 70 kV, 360 mAs) with an isotropic resolution of 31 \textmu m and with a clinical QCT system (Siemens Somatom 64, 120 kV, 100 mAs, kernel B40s) with an in-plane resolution of 0.2 mm and a slice spacing of 1 mm.
	Both scans where calibrated and the QCT scan was resampled and rigidly registered to the \textmu CT scan. The resulting rigid transformation matrix was saved for later use.
	We applied our method to the unregistered QCT scan of the embedded vertebra and sampled the resulting surface as above.
	Using the inverse transformation matrix the point cloud was transformed into the coordinate system of the \textmu CT scan.
	To evaluate the accuracy of the cortical surface we first need to identify the ground-truth cortex in the \textmu CT scan.
	We sampled the \textmu CT scan along lines orthogonal to the acquired surface at every sample point, from 5 mm outside to 5 mm inside the volume defined by the surface using $2001$ samples per line ($5$ \textmu m spacing).
	We binarized the profile using a threshold of $500$ mg/cc.
	To fill small cavities in the cortex (see figure \ref{fig:muCTScan}), we applied a morphological closing operation to the binarized signals.
	We defined the periosteal (outer) surface as the first rising edge and the endosteal (inner) surface as the first falling edge.
	The center of the cortex is then the midpoint between the periosteal and the endosteal surface.
	Since the surface estimated by the proposed method is located at the center of the sampled line ($t = 0$), the signed distance to the real cortex center is simply the location of the midpoint.
	
	\begin{figure}[tb]
		\centering
		\includegraphics[width=0.95\textwidth]{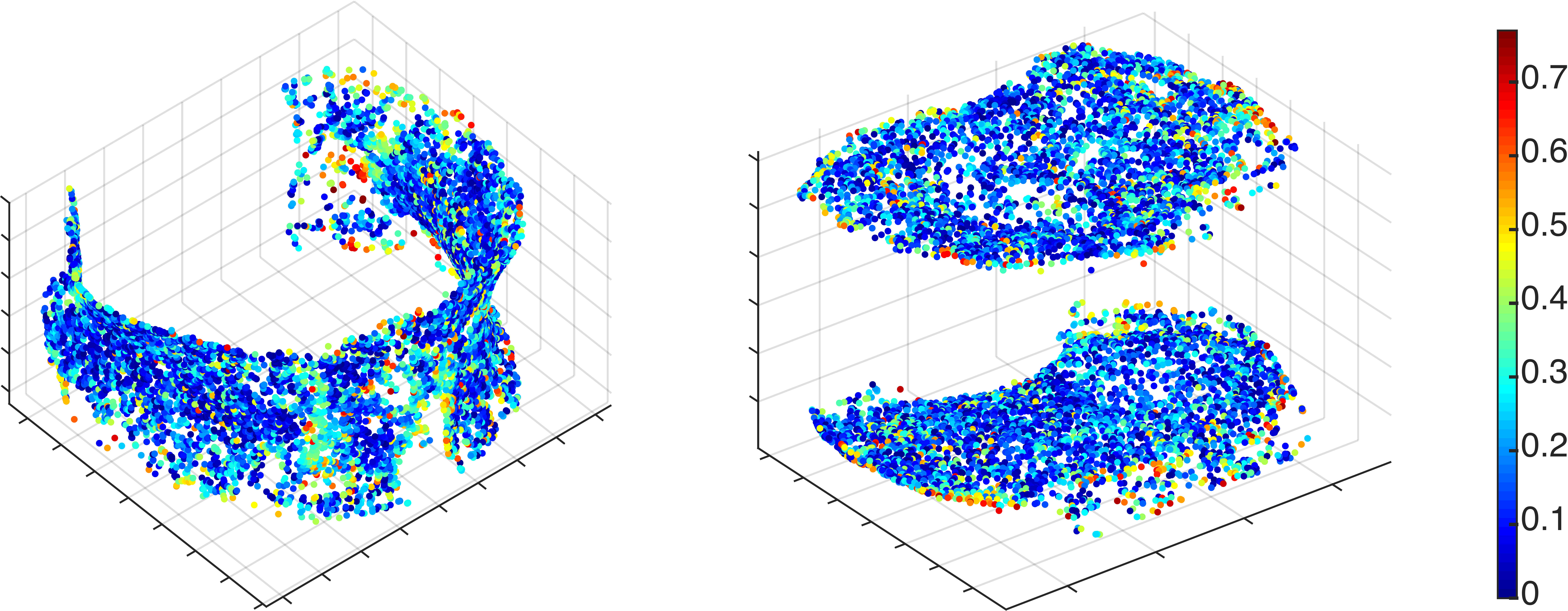}
		\caption{Spatial distribution of absolute distances [mm] to cortex centers of the \textmu CT experiment. Left: vertical cortex, right: endplates. The distances are encoded as colors.}
		\label{fig:MuCTError}
	\end{figure}
	
	We were able to identify the cortex center for $8506$ samples for the vertical cortex and for $8273$ samples for the endplates using this method.
	The average distance to the center of the vertical cortex was $0.0662 \pm 0.2327$ mm (mean $\pm$ standard deviation) and $0.0607 \pm 0.2347$ mm for the endplates.
	The absolute error was below $0.075$ mm for $25$\%, below $0.16$ mm for $50$\% and below $0.28$ mm for $75$\% of the samples.
	Figure \ref{fig:MuCTError} depicts the spatial distribution of the absolute errors on the surface. The highest errors are scattered around the surface, but do not form larger clusters.
	We manually inspected the profiles with the highest errors and found that most of them are located at positions where the real cortex is hard to identify by the thresholding method.
	Figure \ref{fig:muCTScan} shows an example of such a location.
	While the cortex centers identified by the thresholding method vary quite widely, the estimated surface stays smooth between the periosteal and the endosteal surfaces.
	
	\subsection{In-Vivo}
	\label{sec:InVivoExperiments}
	
	For the in-vivo evaluation of the proposed method we used 100 clinical QCT scans (Phillips Brilliance 16, 120 kV, 90-420 mAs) of the first lumbar vertebra from an osteoporosis study with an in-plane resolution of 0.7 mm and a slice thickness of 0.8 mm (increment 0.4 mm).
	The scans were calibrated and segmented by an expert operator.
	In the segmentation process, a triangle mesh representing the apparent cortical ridge surface was exported.
	We used those meshes as a reference to evaluate the applicability of the proposed method to in-vivo data.
	However, since the manually obtained meshes do not identify the center of the cortex but the apparent cortical ridge, a bias is unavoidable.
	After applying the proposed method to the in-vivio scans, we did the same random sampling as in the ex-vivo case and afterwards computed for each sample the distance to the reference mesh using \cite{cignoni1998metro}.
	
	For the vertical cortex we acquired $98000$ samples ($980$ samples per patient) with an average distance of $0.25 \pm 0.52$ mm (mean $\pm$ standard deviation). For the endplates we acquired $101600$ samples ($1016$ samples per patient) with an average distance of $0.19 \pm 0.56$ mm.
	By visual inspection of the spatial distribution of absolute errors, we found that the majority of high errors are located in the transition area between the vertical cortex and the endplates.
	
\begin{figure}[bt]
	\centering
		\includegraphics[width=0.95\textwidth]{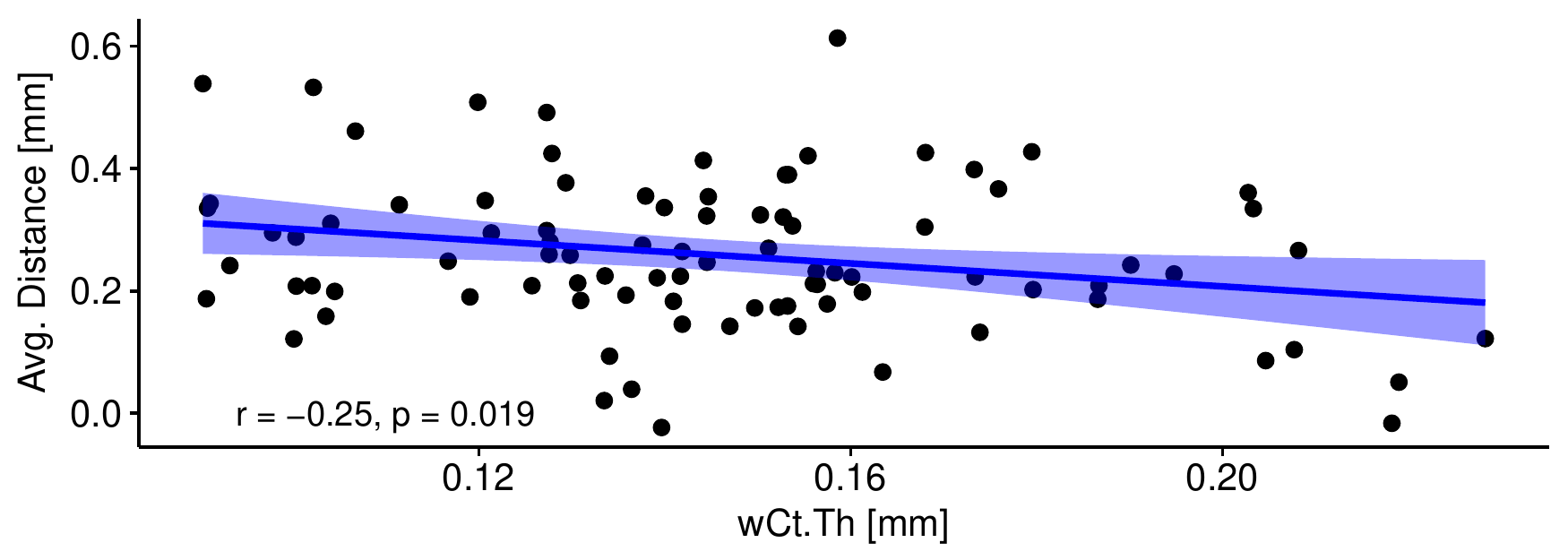}
  		\caption{Scatterplot of the average distance of the estimated to the manual obtained mesh by weighted cortical thickness (wCt.Th). The regression line has a significant negative slope of $-0.94$ ($p = 0.019$) indicating a negative linear relationship. The confidence interval is depicted as the shaded blue region.}
  		\label{fig:BiasByCtTh}
	\end{figure}

	To check if the bias could be explained by the displacement between ridge and cortex center (cf. figure \ref{fig:CortexCenterShift}), we evaluated the 99\% confidence intervals of $\zeta$ \eqref{Eq:ConditionalDensity} for a cortical thickness of $0.3$ mm and for priors $Z_0$ and $Z_2$ estimated from the 100 in-vivo scans.
	For an angle of $\theta = 90^{\circ}$, the displacement between the cortical ridge and the cortex center lies between $0.1$ mm and $0.3$ mm and for an angle of $\theta = 0^{\circ}$ between $0.09$ mm and $2.3$ mm.
	This could explain the observed bias, but for the in-vivo case it cannot be directly verified.
	To substantiate our hypothesis, we did a statistical analysis based on measurements obtained using the original manual segmentation of the 100 in-vivo scans.
	There were no significant correlations between the bias and the integral BMD, cortical thickness, the total volume or the ratio of cortical to trabecular BMD.
	However, the cortical thickness measure that was used is based on voxel distances and therefore it is not very expressive.
	Therefore, we also investigated the density weighted cortical thickness (cortical thickness multiplied by cortical BMD divided by $1200$ mg/cc).
	We found a significant negative linear relationship between the bias and density weighted cortical thickness (wCt.Th). Figure \ref{fig:BiasByCtTh} shows a scatter plot with the regression line.
	There is a significant negative slope of $-0.94$ ($p = 0.019$).
	Therefore, if the cortex gets thicker, the bias gets lower. This is consistent with the cortex center shift as depicted in figure \ref{fig:CortexCenterShift}: a thicker cortex is equivalent to a higher resolution resulting in a better agreement between the cortical ridge and the cortex center.
	
	\section{Conclusion}
	\label{sec:Conclusion}
	
	We presented an analysis by synthesis approach for automatic vertebral shape identification in clinical QCT.
	The foundation of our method is a statistical model of bone which is convolved with the in-plane and out-of-plane PSF of spiral CT resulting in a statistical measurement model.
	We use an as-rigid-as-possible deformation scheme to find the surface of the center of cortical bone that maximizes the posterior likelihood of the model given the image.
	Since the posterior likelihood is analytically intractable, we propose an alternating optimization scheme to find an approximate solution using an efficient data parallel process.
	
	The evaluation of the proposed method using a clinical QCT scan of the ESP and a \textmu CT scan of an embedded vertebra demonstrates its high sub-voxel accuracy.
	The applicability to in-vivo data was shown by comparing the estimated surfaces to manual annotations of 100 in-vivo QCT scans from an osteoporosis study.
	The remaining bias to the manually obtained meshes might be explained by the displacement of the apparent cortical ridge to the cortex center due to low spatial resolution. We substantiated this hypothesis by a statistical correlation analysis.
	
	We think the proposed method is a good starting point for further assessment of cortical bone markers.
	The estimation of the (local) cortical thickness and the cortical BMD should be possible by maximizing the posterior probability of the respective parameters, like we did for the displacements.
	We may also note that, although we presented our method here for vertebral shape identification, it is not limited to vertebrae.
	Using a different template mesh the method can also be applied to other bones or bone parts, e.g. proximal femur.
	
	There are of course limitations of our method: it fails to accurately estimate the cortex center at the transition area between the vertical cortex and the endplates.
	In these areas the assumptions we make for our model are not fulfilled.
	If two cortices are close together, both cortices appear in a single profile, leading to a low likelihood for both cortex positions.
	However, for clinical assessment, those transition areas can be easily excluded.
	We also note that we did not evaluate the precision of our method, yet.
	Since the method is deterministic, it produces the same result for multiple runs on the same input image, but
	the determination of the precision using multiple re-located scans of the same object is still to be done.
	
	To promote the comparability of QCT analysis algorithms we are making our software publicly available\footnote{\url{https://github.com/ithron/CortidQCT}} under open source license.
	
	\subsubsection{Acknowledgments.}
	This work was part of the Diagnostik Bilanz Study which is part of the BioAsset project.
	BioAsset is funded by a grant of the Bundesministerium f\"ur Bildung und Forschung (BMBF), Germany, F\"orderkennzeichen 01EC1005.
	This work was also supported by the German Research Foundation, DFG, No. KO2044/9-1. 

	\bibliographystyle{splncs04}
	\bibliography{references}

\end{document}